\documentclass[twocolumn,amsmath,amssymb,superscriptaddress, prb]{revtex4}
\usepackage[draft]{hyperref}
\usepackage{amsmath}
\usepackage{amssymb}
\usepackage{epsf}
\usepackage{graphicx}
\usepackage{tabulary}
\usepackage[version=3]{mhchem}
\usepackage{color}
\usepackage{upgreek}

\begin{document}
\title{Bismuth Ferrite Dielectric Nanoparticles Excited at Telecom Wavelengths as  Multicolor Sources by Second, Third, and Fourth Harmonic Generation}
\author{Jeremy Riporto} \affiliation{Universit\'e
Savoie Mont Blanc, SYMME, F-74000 Annecy, France} \affiliation{GAP-Biophotonics, Universit\'e de Gen\`eve, 22 chemin de Pinchat, 1211 Gen\`eve 4, Switzerland}
\author{Alexis Demierre} \author{C\'edric Schmidt} \author{Gabriel Campargue} \author{Vasyl Kilin} \affiliation{GAP-Biophotonics, Universit\'e de Gen\`eve, 22 chemin de Pinchat, 1211 Gen\`eve 4, Switzerland} 
 \author{Tadas Balciunas}\affiliation{Photonics Institute, TU Wien, Gusshausstrasse 27/E387, 1040, Vienna, Austria}
\author{Mathias Urbain}\affiliation{Universit\'e
Savoie Mont Blanc, SYMME, F-74000 Annecy, France}

 \author{Andrius Baltuska}\affiliation{Photonics Institute, TU Wien, Gusshausstrasse 27/E387, 1040, Vienna, Austria}

\author{Ronan Le Dantec}\affiliation{Universit\'e
Savoie Mont Blanc, SYMME, F-74000 Annecy, France} 
\author{Jean-Pierre Wolf} \affiliation{GAP-Biophotonics, Universit\'e de Gen\`eve, 22 chemin de Pinchat, 1211 Gen\`eve 4, Switzerland} 
\author{Yannick Mugnier}\affiliation{Universit\'e
Savoie Mont Blanc, SYMME, F-74000 Annecy, France} \email{Yannick.Mugnier@univ-smb.fr} 
\author{Luigi Bonacina} \affiliation{GAP-Biophotonics, Universit\'e de Gen\`eve, 22 chemin de Pinchat, 1211 Gen\`eve 4, Switzerland} \email{luigi.bonacina@unige.ch}

%

\begin{abstract}
We demonstrate the simultaneous generation of second, third, and fourth harmonic from a single dielectric Bismuth Ferrite nanoparticle excited by a telecom fiber laser at 1560 nm. We first characterize the signals associated with different nonlinear orders in terms of spectrum, excitation intensity dependence,  and relative signal strengths. Successively, on the basis of the polarization-resolved emission curves of the three harmonics, we discuss the interplay of susceptibility tensor components at the different orders and we show how polarization can be used as an optical handle to control the relative frequency conversion properties.


\end{abstract}

\maketitle

\section*{Introduction}
The generation and control of nonlinear parametric signals at the nanoscale is paving the way to novel applications in imaging, sensing, optoelectronics. To date, most of the research efforts have been concentrated on noble metal nanoparticles and nanostructures\cite{Kauranen2012} with a focus on their second ($\chi^{(2)}$)\cite{Dadap1999, Butet2010} and third order ($\chi^{(3)}$)\cite{Lippitz2005, Danckwerts2007} response. Some notable exceptions include the nonlinear harmonic generation by semiconductor nanoparticles\cite{Jacobsohn2000, BarElli2015}, two dimensional materials\cite{Dean2009, Hong2013,Kumar2013, Karvonen2015}, and noncentrosymmetric metal oxide nanoparticles (Harmonic NanoParticles, HNPs). Dielectric HNPs are attracting growing interest because of their extremely high nonlinear coefficients,\cite{Rogov2015} and robustness of their nonlinear response  which  - contrary to noble metal particles  -  is primarily associated with their bulk properties and negligibly affected by surface phenomena.\cite{Staedler2012, Kim2013} Moreover, the sub-wavelength dimensions of HNPs lift the spectral limitations imposed by phase-matching conditions in bulk nonlinear crystals, enabling wide tunability of excitation light and  emission of multiple signals at once. Some research groups are  working on the efficiency enhancement of the optical properties by engineering hybrid systems based on a HNP-core and a plasmonic-shell tailored for specific spectral resonances.\cite{Pu2010, Richter2014}   

Recently, we have demonstrated the simultaneous acquisition of Second and Third Harmonic Generation (SHG, THG) by bare individual perovskite Bismuth Ferrite (BiFeO$_3$, BFO) HNPs.\cite{Schmidt2016} We showed that the coincident acquisition of both harmonics can strongly benefit to imaging selectivity in optically congested environments\cite{Rogov2015b} for applications including cell-tracking over long time in tissues.\cite{Dubreil2017} Besides harmonic generation, one can expect that high $\chi^{(n)}$ values by HNPs can be exploited for disposing of localized sources of long wavelength radiation by optical rectification or for generating nonclassical states of light, in analogy to what has been demonstrated using other kinds of nanostructures.\cite{Polyushkin2011, Grice1997, Dot2012, Wu1986}    In this respect, the possibility of working efficiently at telecom wavelengths (1.5 $\upmu$m)  undeniably constitutes  an asset for a future integration of HNPs as frequency conversion elements and all-optical logic operators\cite{Puddu2004} in photonics circuits. 

In this work, we  demonstrate that second, third, and fourth harmonic (FHG) emitted by an individual BFO HNP upon excitation at 1560 nm by an Erbium-doped fiber oscillator can be efficiently detected. Moreover, we show how the polarization control of excitation light allows tuning the relative intensities of the three harmonics. The simultaneous acquisition of three harmonics from the same individual nanoparticle is - to our best knowledge - a \textit{unicum} to date and, besides all the applications we mentioned,  HNPs might assume the role of model system for the study of the interplay among multiple-harmonics and high harmonic generation in solids.\cite{Ghimire2011, Vampa2015, Luu2016, Saltiel2004}   

\section*{Results and Discussion}
\begin{figure}
\includegraphics[width=8cm, keepaspectratio]{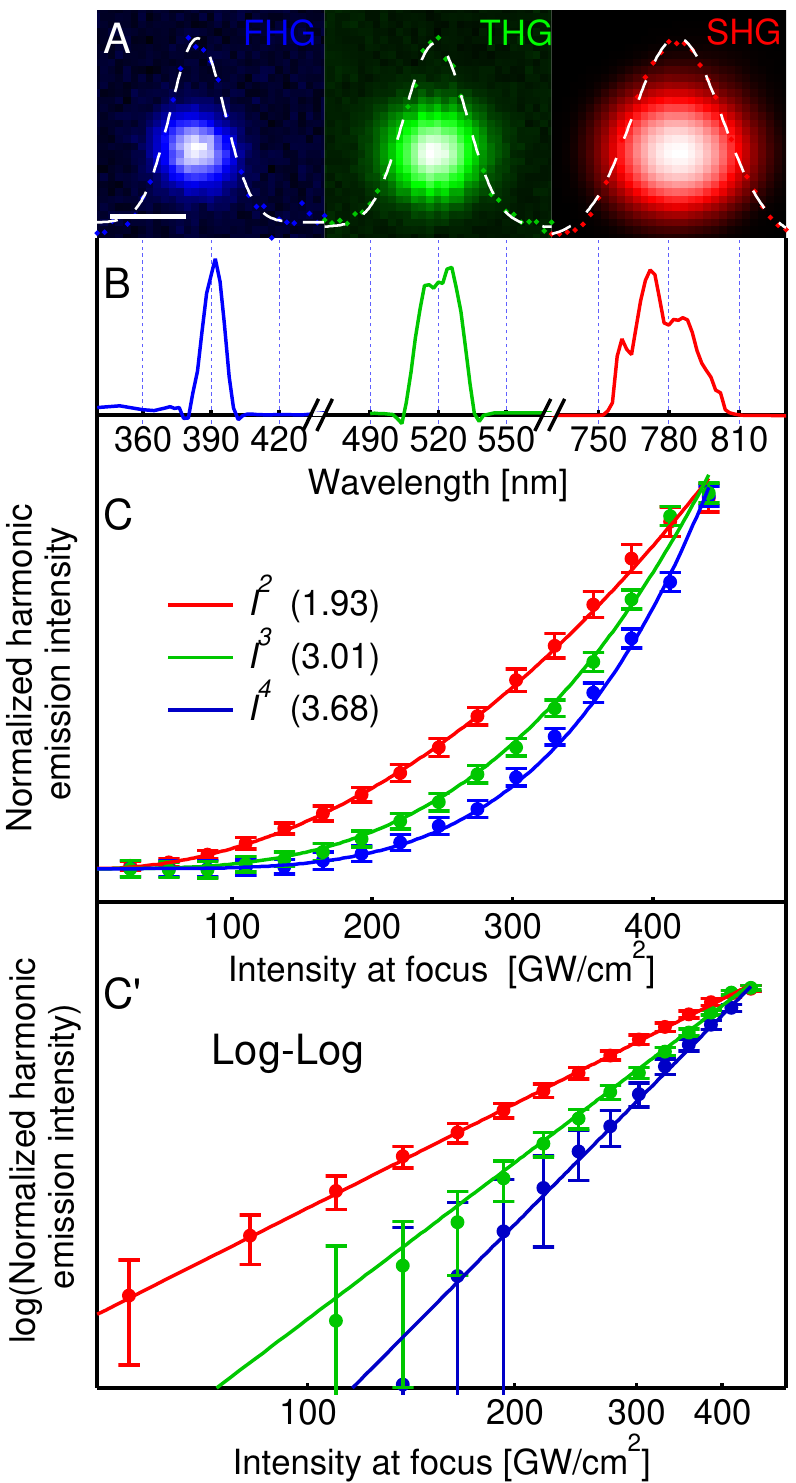}
\caption{\textbf{A.} Images taken at the different harmonics of a single isolated BFO HNP.  The intensity profile obtained at each harmonic (colored dots) is fitted by a Gaussian function (dashed lines). Scale bar 500 nm. \textbf{B.}  Normalized harmonic spectra. The interval between two dashed vertical lines corresponds to 30 nm. \textbf{C.} Normalized power dependence of the intensities of the three harmonic emissions. The  continuous lines represent the nominal  $I^n$ traces with $n=2,~3,~4$.  The optimal $n$ values obtained by fitting  the experimental traces are reported in parentheses (fits not shown).  \textbf{C'.} Log-Log representation of the data in panel C.} \label{Fig:1}
\end{figure}

The starting evidence motivating this work is the observation that single BFO HNPs deposited on a substrate in the focus of the laser emit simultaneously at the three harmonics as from the images in Fig.~\ref{Fig:1}A. The heat-maps colors are red for SHG (780 nm), green for THG (520 nm), and blue for FHG (390 nm). In the following, we first present a thorough assessment demonstrating by independent experimental observables [i) image spot size, ii) spectrum, and iii) excitation intensity dependence] that the three emission are genuinely associated with different nonlinear orders. Successively, we discuss the polarization-resolved emissions at the different orders which shed light on the tensorial properties of the nonlinear susceptibilities and could prospectively be exploited for selective frequency up-conversion from short-wave infrared to the visible. 

i) The Gaussian fits to the diameters of the particle images in Fig.~\ref{Fig:1}A indicate that the FWHM decreases with increasing nonlinearity, as one  expects for a diffraction limited object smaller than the point spread function (PSF) at the highest order. The observed widths of the PSF  range from 673  nm for SHG, to 486 nm for THG and 420 nm for FHG.  The average dimensions of the HNPs ($\approx$ 100 nm, Fig.~S1) remain therefore out of reach at all orders.  In the Supplementary Material,  we further comment these results in the context of the imaging properties of the set-up.  ii)  In Fig.~\ref{Fig:1}B, we provide the normalized harmonic spectra detected in the forward direction. In the wavelength domain, one expects the width of the emission to scale as $\propto \frac{1}{n \sqrt{n}}$, with $n$ being the nonlinear order. This formula is derived for Gaussian pulses in the time domain.\cite{Ehmke2015} Therefore, to apply this estimation to our traces stemming from a structured spectral profile at the excitation wavelength, we proceeded by visually determining the broadest Gaussian curves supported by the excitation and by each harmonic spectrum (Fig. S2). This way,  we obtained  widths for the different harmonics within 10\% deviation from the theoretical estimation. This procedure, although involving approximations, points to a rather complete upconversion of the frequencies in the fundamental spectrum and it is consistent with the fact that BFO HNPs are smaller than the coherence lengths for each nonlinear order, $l_c^{(n)}$. By using the optical constants of BFO derived by Kumar \textit{et al.}\cite{Kumar2008}   and applying a calculation including the effect of Gouy phase (see Eq. S1) we obtain $l_c^{(n)}$ values in the forward direction spanning from  1 $\upmu$m  for $n=2$ to 0.325 and  0.23 $\upmu$m   for  $n=3$ and 4, respectively.\cite{Cheng2002} In our calculations, $l_c^{(n)}$ deem  larger than  HNPs typical dimensions. This implies that no destructive interference takes place within the particle volume. iii) To complete this preliminary assessment of multiorder response,  in Fig.~\ref{Fig:1}C, we present the harmonic signal strength as a function of the laser intensity at the sample, $I$. Note that for this comparison the signals are normalized at the maximal laser intensity of the series, which corresponds to 440 GW/cm$^2$. As discussed in the next section, in absolute terms the THG is by far the most intense under these excitation/detection conditions: roughly 2 orders of magnitude stronger than SHG and 4 orders stronger than FHG.  In the image, the nominal fitting curves (\textit{i.e.},  $ I^{n}$, $n=2,~3,~4$) are plotted as continuous lines. One can appreciate their fairly good agreement with the experimental data.  In the  legend, we report the optimal  values for the exponent $n$ obtained letting this parameter free to vary in the fitting procedure. The retrieved values  are all within 10\% deviation from the theoretical values. In Fig.~\ref{Fig:1}C', the  data and fitting curves from panel C are provided in log-log representation.   Altogether these results  obtained by independent  measurements (nonlinear PSF, harmonic spectra,  intensity dependence)  support the association of the signals from single HNPs with three different harmonics: SHG, THG, and FHG. 

\subsection*{Relative intensities of harmonic orders.}

\begin{figure}
\includegraphics[width=7cm, keepaspectratio]{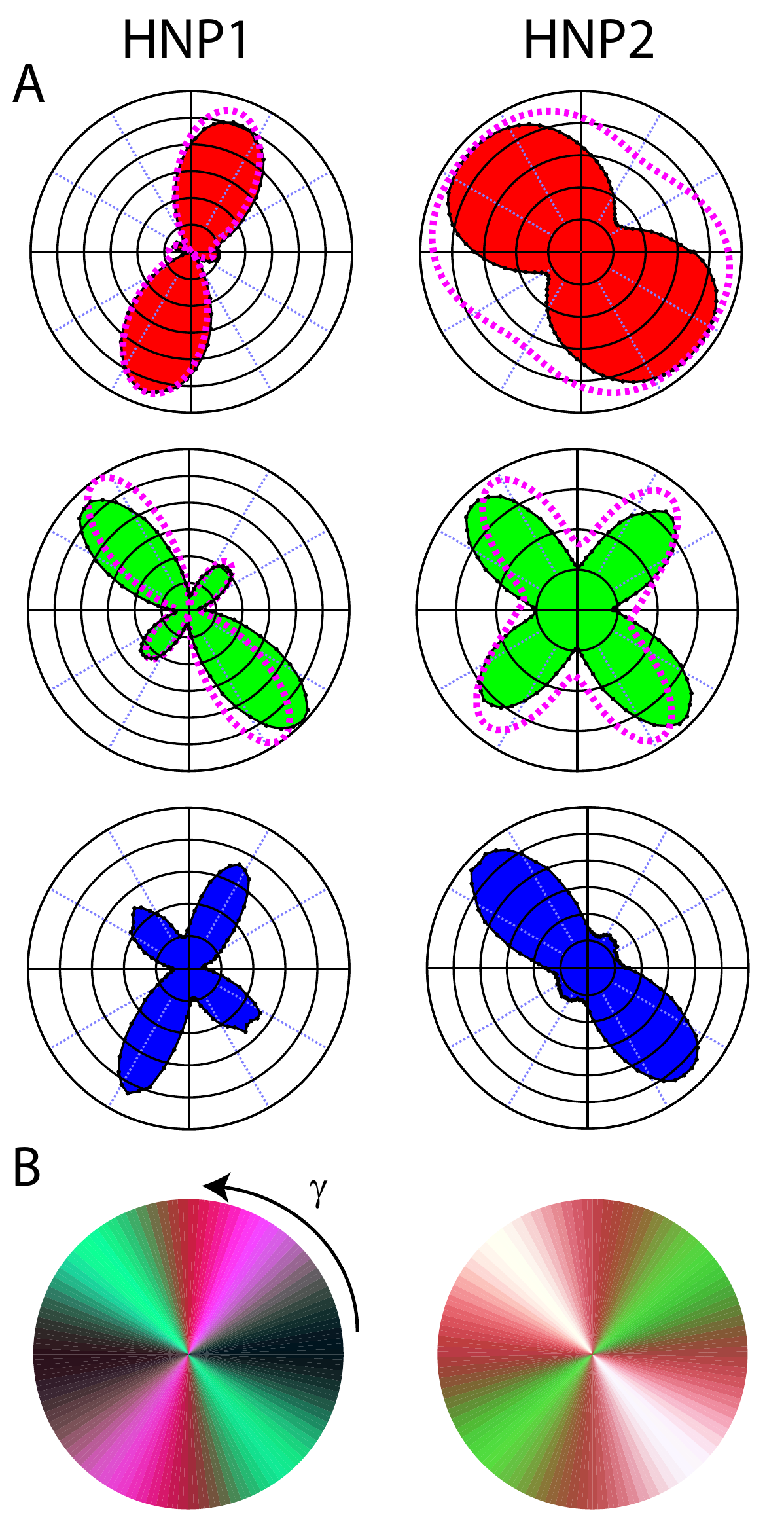}
\caption{Polarization sensitive response. \textbf{A.} The shaded regions represent the emission intensity at the different harmonic orders from two isolated BFO HNPs as a function of the polarization angle $\gamma$. The dashed purple line are  fits to the traces obtained using the $\chi^{(2)}$ and $\chi^{(3)}$ tensors reported in Schmidt \textit{et al.}\cite{Schmidt2016} for 1064 nm excitation.  The corresponding  Euler angles sets  $(\phi,~\theta,~\psi)$ we retrieved are (78$^{\circ}$, 38$^{\circ}$, 314$^{\circ}$) for HNP1 and (68$^{\circ}$, 13$^{\circ}$, 77$^{\circ}$) for HNP2, respectively. \textbf{B.} Alternative representation to highlight the selective frequency generation  obtained by varying the polarization angle $\gamma$. The images are created by adding as RGB components the \textit{normalized}  SHG, THG, and FHG polarization-resolved traces in  A.} \label{Fig:2}
\end{figure}

A natural question arises concerning the relative intensities of the three emissions, as one would normally expect a major decrease in signal strength with increasing nonlinear rank, provided that the symmetry requirements (\textit{i.e.}, noncentrosymmetricity) are fulfilled for the generation of even orders (SHG, FHG,...). Clearly, one should also take into proper account the different intensity dependence exhibited by signals associated with each $\chi^{(n)}$: higher excitation intensity is expected to favour higher orders as the ratio of two successive harmonics scales as $\propto \frac{1}{I}$. \cite{Dai2014} In a previous Hyper Rayleigh Scattering experiment at 1064 nm, we determined that SHG/THG$\simeq$40 for 11 GW/cm$^2$ excitation.\cite{Schmidt2016} Therefore, we could expect this ratio to be here $\leqslant$1 when working at 53-fold larger intensity, \textit{viz.} 590 GW/cm$^2$ (280 pJ/pulse). However, we observe a surprisingly smaller SHG/THG value,  of the order of 10$^{-2}$-$10^{-3}$. Sample resonances  play an important role in determining the value of harmonic ratios:\cite{Kuznetsov2016} a resonance was reported at 504 nm for BFO 25 nm thin films\cite{Kumar2008}  supporting the efficient generation of THG observed here at 1560 nm excitation (a green spot is visible by naked eyed on small aggregates).  This close-to-resonance condition can also help explaining the very high second order susceptibility reported for BFO HNPs excited at 1064 nm, which was estimated to 160 pm/V.\cite{Schwung2014} The FHG/THG ratio is, on the other hand, of the order of 10$^{-4}$. Being aware that, among all techniques, the values extracted by microscopy present the largest uncertainty because they imply averaging the response of individual particles (10 in the present case, Fig. S3) with different spatial orientations modulating their harmonic ratios, we complemented these measurements with  additional ones performed on  pellets of compressed BFO HNPs (Fig. S4A). These measurements were carried out at 1 TW/cm$^2$ using a $\upmu$J laser system, averaging the response of a large ensemble of randomly oriented particles over an elliptic area of 60$\times$120 $\upmu$m$^2$. By this approach, we obtained SHG/THG$\approx$20 (Fig. S4B) while the FHG/THG is $\leq$ 10$^{-4}$. Such a large discrepancy among the outcomes of the two methods,  in particular for the SHG/THG ratio, is not fully clear. On one hand, the presence of aggregates in the pellets with dimensions exceeding the coherent length  of BFO can affect the signal in an uneven way throughout the spectral domain. Moreover, the comparison can be also undermined by the difficulty to find a meaningful definition of peak intensity encompassing both large particles ensembles and isolated objects substantially smaller than the focal spot size. Finally, the difference  observed can be ascribed to the critical dependence  of coherent signals generated by individual nanostructures on experimental settings (\textit{e.g.}, N.A. and collection angle). This last aspect has been subject of multiple theoretical studies in the plasmonic community based on different approaches (method of moments,\cite{Araujo2012} finite elements,\cite{Poutrina2013} hydrodynamic model\cite{Sipe1980, Ciraci2013}). Recently, the hydrodynamic approach has been applied to calculate the SHG and THG angular radiation patterns simultaneously emitted by individual plasmonic nanoparticles, which specifically highlights this sensitivity to detection parameters showing rather different angular emission patterns at the two harmonics.\cite{Ginzburg2014} We believe that only a rigorous extension to higher harmonics of Hyper Rayleigh Scattering on colloidal suspensions can provide reliable values for the material.\cite{Joulaud2013}

The comparatively high conversion efficiencies at the third order we observe by both approaches  for a noncentrosymmetric  material  displaying very high quadratic nonlinearity such as BFO, can  also be potentially  ascribed to the presence of multi-step (cascading) processes involving a succession of purely $\chi^{(2)}$ phenomena: SHG and sum frequency mixing.\cite{Saltiel2004} In this case, THG would result from $\omega+\omega=2\omega$ and $\omega+2\omega=3\omega$,\cite{Bosshard2000} whereas FHG from $\omega+\omega=2\omega$ and $2\omega+2\omega=4\omega$ or, alternatively, from  $\omega+\omega=2\omega$ followed by  $2\omega+\omega=3\omega$ and  $3\omega+\omega=4\omega$.\cite{Ivanov2005} It is tempting to attribute the comparatively low emission at $2\omega$ to a depletion of this frequency used as intermediate field for generating $3\omega$, however discerning multi-step from direct higher order nonlinear processes is a complex task, in particular for nanoparticles as the absence of macroscopic propagation excludes discrimination methods based on phase-matching criteria.\cite{Bosshard2000} The use of HNPs with controlled size and narrow size distribution or epitaxial thin films of variable thickness could help elucidating this aspect in a future series of experiments. 

\subsection*{Polarization properties.}
In Fig.~\ref{Fig:2}A, we introduce the results on polarization dependence for two sub-diffraction limited and  isolated particles: HNP1 and HNP2.  The shaded regions  display the intensity of the harmonic emission detected as a function of the polarization angle of the excitation laser,  $\gamma$. Note that differently from other works,\cite{Brasselet2004, Bonacina2007, Schmidt2016} in this case no polarization analyser was set in the detection arm. The differences between the response of the two HNPs are associated with the different spatial orientations of their crystal axis with respect to the laboratory frame (Euler angles $\phi, \theta, \psi$ in Fig.~\ref{Fig:3}). The simple inspection of the polarization resolved traces can provide precious information and it deems useful to discard from the analysis eventual polycrystalline aggregates.\cite{Brasselet2004, Bonacina2007} In general, the SHG traces possess a structure characterized by two dominant lobes in  agreement with our previous observations.\cite{Schmidt2016} For THG and FHG the side lobes become more prominent. Interestingly, the orientation of the main lobes is mostly maintained among the even orders (SHG, FHG) while for THG it seems that other tensor elements become predominant with  major changes in  orientation and symmetry. In our previous study,\cite{Schmidt2016} starting from a known $\chi^{(2)}$ tensor,\cite{Kumar2008} we fitted the orientation of several BFO particles and then used the retrieved Euler angles  to determine the unknown $\chi^{(3)}$ tensor elements by simultaneously fitting the THG response of several HNPs. Here, we use these tensor values for $\chi^{(2)}$ and $\chi^{(3)}$  to fit the  SHG and THG traces and obtain the Euler angles of each particle. The fits are reported as purple dashed lines on the data and the angle sets for HNP1 and HNP2 provided in the figure caption. Although the fits correctly capture the main features of the polarization curves (main lobes angles, presence of orthogonal lobes), one should be aware that this procedure implies several approximations  and the result should be considered qualitative in nature and primarily intended to support the fact that the BFO point group (3\textit{m}) is compatible with the observed traces. In particular, the tensors we  apply are derived at 1064 nm and not at 1560 nm.  Note also that we could not readily extend this approach to  $\chi^{(4)}$  because the number of independent elements of this tensor prevents the retrieval of a reliable outcome. Finally, we highlight that the possible presence of competing multi-step $\chi^{(2)}$ processes would  undermine the general validity of this description, which would remain however an effective tool for predicting the polarization dependent response of BFO HNPs even in presence of concurrent direct and cascaded generation.

\subsection*{Polarization-based control of relative harmonic intensities.}
The response of the two randomly oriented HNPs suggests that the choice of the excitation polarization, even in absence of any detection analyser, can be used to modulate the \textit{relative} intensities of the three emissions for a given laser polarization angle $\gamma$. In Fig.~\ref{Fig:2}B, we graphically emphasize this procedure showing the total emission  obtained by adding the \textit{normalized} polarization dependent harmonic components displayed by HNP1 and HNP2. This alternative representation shows how, for a given HNP orientation, specific values of $\gamma$ are associated with strong simultaneous SH, TH, FH emission (white regions),  with individual harmonics (red, green, blue in our representation) and combination thereof (purple, pink...) or low emission (dark regions). We speculate that this approach could be adapted to precisely oriented BFO HNPs and  thin films with thickness smaller than the shortest coherence length to provide polarization-controlled frequency converters from the telecom region over the  visible spectrum. Engineered hybrid structures composed by HNPs with a plasmonic shell of tailored thickness or, alternatively, the choice of materials with  tailored resonances,\cite{Kuznetsov2016} could also be a  way to mitigate the large conversion efficiency differences at the three  harmonic orders for defined applications.\cite{Pu2010, Richter2014}  Alternatively, one could shape the excitation geometry to control to some extent the angular emission pattern at the different orders.\cite{Carletti2016}

\section*{Conclusions}

In conclusion, we have reported what,  to our best knowledge, is the first demonstration of simultaneous acquisition of three harmonic frequencies generated by an isolated nanoparticle. Notably, our experiment is performed using a pJ fiber laser at telecom wavelength,  which holds great promise for implementing  dielectric nonlinear nanophotonics\cite{Smirnova2016} in optoelectronic circuitry.  Considered the novelty of our observation, we first thoroughly assessed the spectral and imaging properties and the intensity dependence of the emissions to ensure that they are genuinely associated with frequency conversion  by $\chi^{(n)}$  ($n=2,~3,~4$) or cascaded $\chi^{(2)}$ processes. The relative intensities of the three harmonics have been critically discussed highlighting the sensitivity of this parameter to the measurement method. All estimations point to  high generation efficiency for THG, likely because of the presence of electronic resonances in the spectrum. Finally, we  have discussed the excitation-polarization dependence of the particle emission, demonstrating that this approach  opens the way to  directly  investigating the interplay among nonlinear susceptibility tensors elements at different orders and  modulating the relative strengths of three color components (red, green, violet) for photonics applications.

\section*{Methods}   
\begin{figure}
\includegraphics[width=8cm, keepaspectratio]{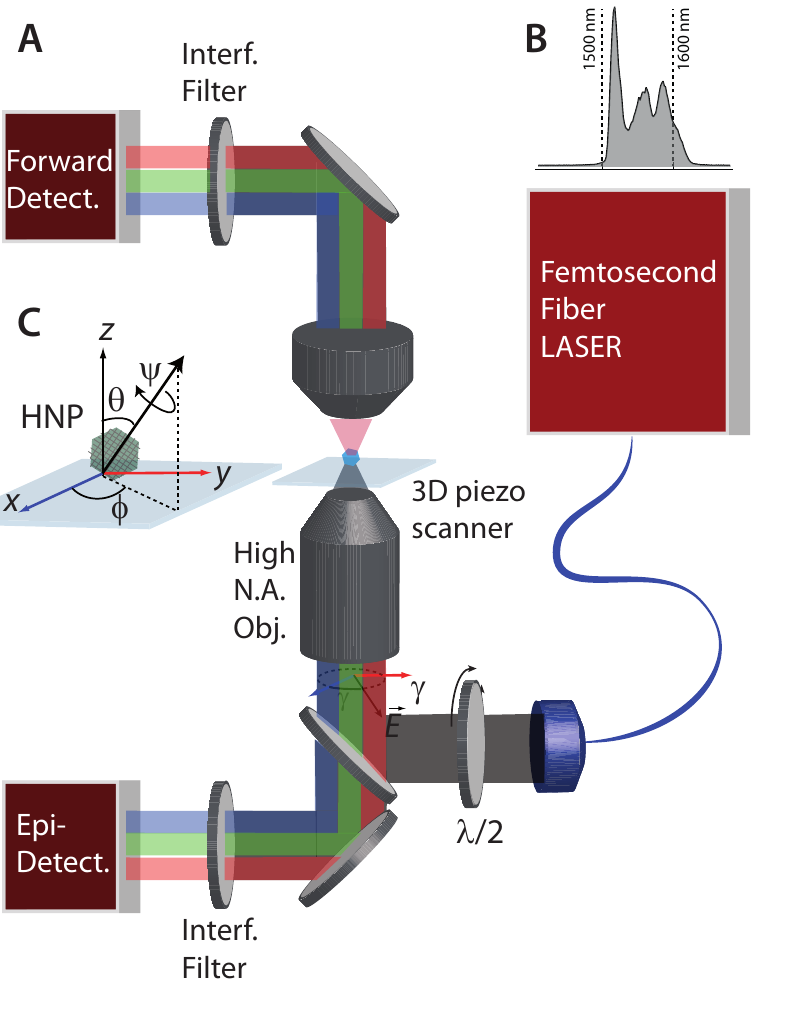}
\caption{\textbf{A.} Schematics of the experimental set-up. Not shown: a spectrometer can be inserted in the forward detection arm and intereference filters removed for acquiring spectrally resolved traces. \textbf{B.} Measured laser spectrum.  \textbf{C.}  Euler angles defining the HNP crystal axis orientation with respect to the laboratory frame.} \label{Fig:3}
\end{figure}
BFO nanoparticles synthesized by the company FEE GmbH (Idar-Oberstein, Germany) were obtained as a water stabilized colloidal suspension from the company  TIBIO (Comano, Switzerland) under a research agreement. The average size is estimated to $\approx$ 100 nm by dynamic light scattering (DLS) and transmission electron microscopy (Fig. S1).  For imaging,  a drop of BFO suspension  is cast onto a microscope substrate and the solvent let evaporating. 

As reported in Fig.~\ref{Fig:3}, the light source of the set-up is a Telecom femtosecond fiber laser at 1560 nm with a repetition rate of 100 MHz  and 100 mW average power (\textit{T-Light FC}, Menlo Systems). Pulses are compressed down to 90 fs by an optical fiber connected to the laser output. At the fiber output, the beam is collimated in the free space and expanded to a diameter of 6 mm.  For polarization resolved studies, the linear polarization of the laser is rotated by a $\lambda/2$ plate  mounted on a motorized rotation stage. In the case of power dependence measurements, the laser energy is continuously modulated through the succession of  a $\lambda/2$  plate and a polarisation analyser.  Afterwards, the beam is reflected by a 45 degrees short-pass filter (Chroma) and focused on a single isolated HNP by a  100$\times$ microscope oil immersion objective (NA 1.3).  The signal generated by the particles can be detected in the backward or forward direction. In the latter case, the collection objective is a 40$\times$ N.A. 0.6  air objective. HNPs are  selected by scanning a $(x,y)$ planar ROI of approximately 20$\times$20 $\upmu$m$^2$ with a piezo-stage and carefully adjusting the $z$ position by maximizing their nonlinear signal.  Both for epi- and forward-detection, narrow bandwidth interference filters are used to select the harmonic spectral region (Thorlabs \textit{FBH780-10} for SHG, \textit{FBH520-40} for THG, \textit{FBH400-40} for FHG  and  Semrock \textit{BrightLine Fluorescence Filter 387/11}  for FHG). Additionally, a scanning spectrometer (\textit{Acton SP2300}, Princeton Instruments, 300 g/mm) is placed in the forward detection arm to acquire spectrally resolved traces. The measurements are obtained using two different Hamamatsu detectors, selected according to their spectral response: \textit{H7732-01} low noise side-on photomultiplier tube  (185 nm to 680 nm), and  \textit{H7421-50} photon counting head with a GaAs photocatode (380 nm to 890 nm). Alternatively, we use a ultra-low-noise single photon counting module (\textit{SPD-A-VISNIR}, Aurea Technology, Besan\c{c}on, France).

\section{Acknowledgements}

 We acknowledge the financial support by Swiss SEFRI (project C15.0041, Multi Harmonic Nanoparticles), by the French-Switzerland Interreg programme (project NANOFIMT), and  by the NCCR Molecular Ultrafast Science and Technology of the Swiss National Science Foundation. This study was performed in the context of the European COST Action MP1302 Nanospectroscopy.

 We are grateful to Dr. Davide Staedler at TIBIO  SA (Comano,  Switzerland) and Dr. Daniel Rytz at FEE GmbH (Idar-Oberstein, Germany) for synthesizing and providing us colloidally stable BFO HNPs,  to Dr. Johann Cussey from Aurea Technology (Besan\c{c}on, France) for providing us the  the single photon counting module and technological support,  and Virginie Monnier (Institut des Nanotechnologies, Lyon) for  the TEM images of BFO HNPs.


\providecommand*{\mcitethebibliography}{\thebibliography}
\csname @ifundefined\endcsname{endmcitethebibliography}
{\let\endmcitethebibliography\endthebibliography}{}

\onecolumngrid
\newpage
\setcounter{section}{0}
\setcounter{figure}{0}
\section{Supplementary Material}
\section{BFO HNPs characterization}
\begin{figure}[h]
\centering
\includegraphics[width=7cm, keepaspectratio]{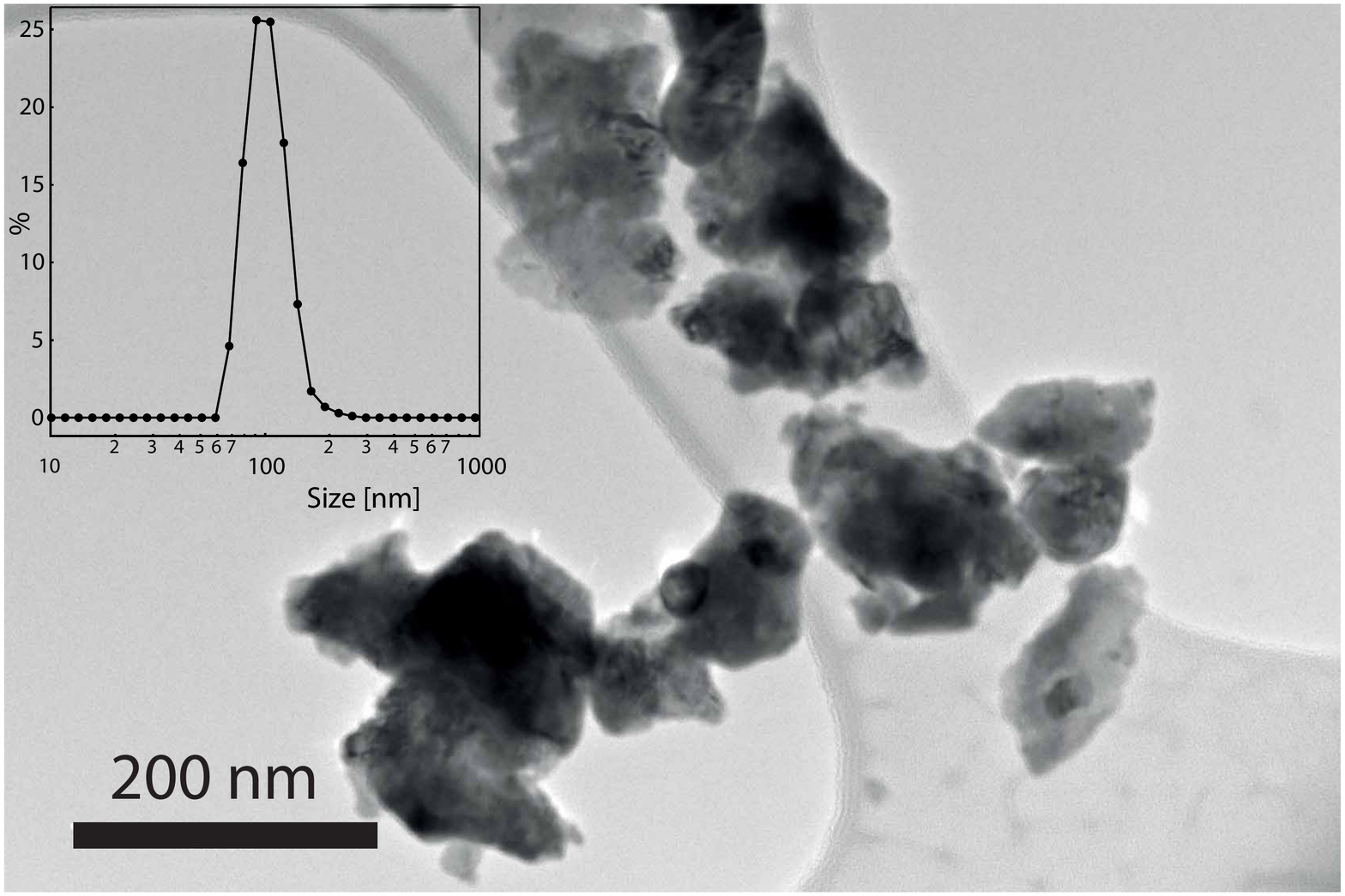}
\caption{TEM image of BFO HNPs and plot of the Dynamic Light Scattering distribution by number.} \label{FigSI:1}
\end{figure}

A detailed description of the synthesis and properties of the nanoparticles used in this work can be found in Schwung \textit{et al.}, \cite{Schwung2014}  TEM and DLS representative data of a sample obtained by this protocol are reported in Fig.~\ref{FigSI:1}.

\section{Width of the Point Spread Function (PSF) at the different harmonic orders}
Taking into account excitation wavelength and  objective numerical aperture, the nominal lateral FWHM of a perfect imaging system under linear excitation should be FWHM$_{linear}^{theo}$=0.51$\lambda$/N.A.=612 nm.\cite{Wilson2011} For the nonlinear case, Zipfel \textit{et al.} provide the following expression for a two-photon excited fluorescence emitter: FWHM$_{2nd~order}^{theo}$=$2\sqrt{\ln{2}}\frac{0.325\lambda}{\sqrt{2} NA^{0.91}}$= 391 nm.\cite{Zipfel2003} These values cannot be applied here because the resolution is expected to be severely reduced by the fact that we are using an high N.A. oil immersion objective intended for the visible region and not for an excitation at 1.5 $\upmu$m. Therefore all aberration corrections and optical elements (comprising the matching medium) are far from optimal. Indeed, we observe an energy reduction of 75\% upon laser transmission through this objective, indicating a poor compatibility at this wavelength. By considering that the resolution should be proportional to $1/ \sqrt{n}$ where $n$ is the nonlinear order, we can readily compute an actual value of $\approx$840 nm for the width of the linear PSF, both by multiplying the FWHM$_{FHG}$ (420 nm) by $\sqrt{4}$ and FWHM$_{THG}$  (486 nm)  by $\sqrt{3}$.  Note that this result supports the fact that we are observing a sub-diffraction limited emitter at two harmonic orders. The same calculation applied to the FWHM$_{SHG}$ (673 nm)  provides a result $\approx$15\% higher. In this series, SHG was epi-detected using the \textit{H7421-50} photon counting  and THG and FHG forward detected by the \textit{H7732-01} low noise side-on photomultiplier tube.  The 15\% discrepancy can very likely be attributed to the deviation from linear response of the former detector in the intensity regime of the measurement. 

Note that the FWHM value of 840 nm was used for the microscopy-based  intensity ratio calculation.

\section{Estimation of the widths of the harmonic spectra}

\begin{figure}[t]
\centering
\includegraphics[width=7cm, keepaspectratio]{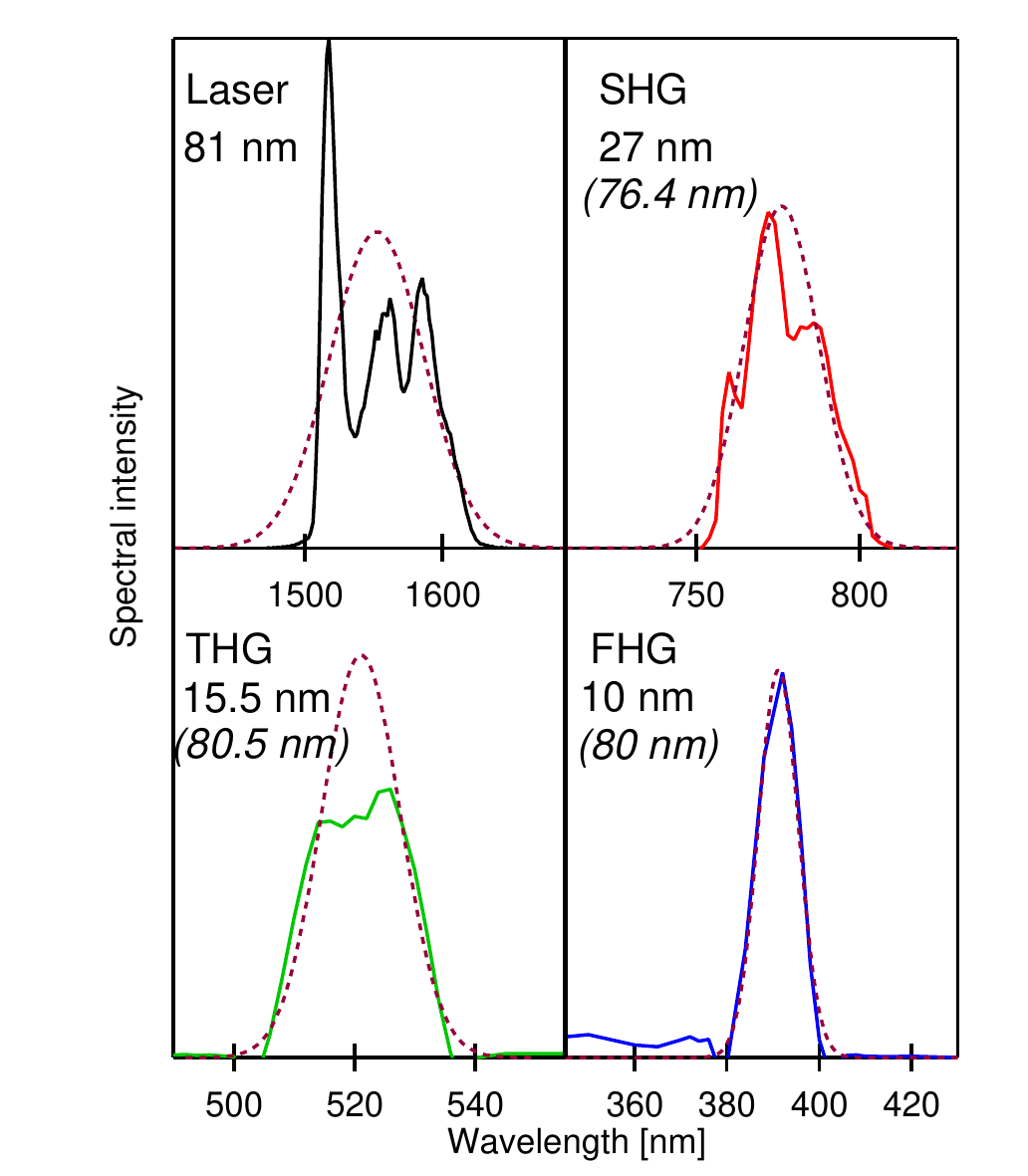}
\caption{Normalized spectra of laser and different harmonics (continuous lines) along with Gaussian curves supported by the spectra (dashed lines).} \label{FigSI:2}
\end{figure}

In Fig.~\ref{FigSI:2}, we report the normalized spectra of the laser and of the three harmonics generated by a single BFO HNP along with Gaussian curves  supported by these spectra and determined by visual inspection.  On the figure we provide the Gaussian FWHM and, in parentheses,  the product FWHM$\cdot n\sqrt{n}$ which should be directly compared with the laser spectrum as discussed in the main text.

\section{Calculation of coherent lengths at different orders}

The coherence length is estimated using 

\begin{equation}
l_c^{(n)}=\frac{\pi}{k(n\omega)-n k(\omega)-n \Delta k_G}
\end{equation}

where $n=2,3,4$ for SHG, THG, and FHG, respectively.  $\Delta k_G$ is the wave vector corresponding to the Gouy-phase shift.  The numerical value of $\Delta k_G$ was estimated at -0.5$\pi/\lambda$ by Cheng and Xie for a 1.4 N.A. objective.\cite{Cheng2002} 


\section{Intensity ratios}

\subsection{Measurements on individual particles}

\begin{figure}
\centering
\includegraphics[width=7cm, keepaspectratio]{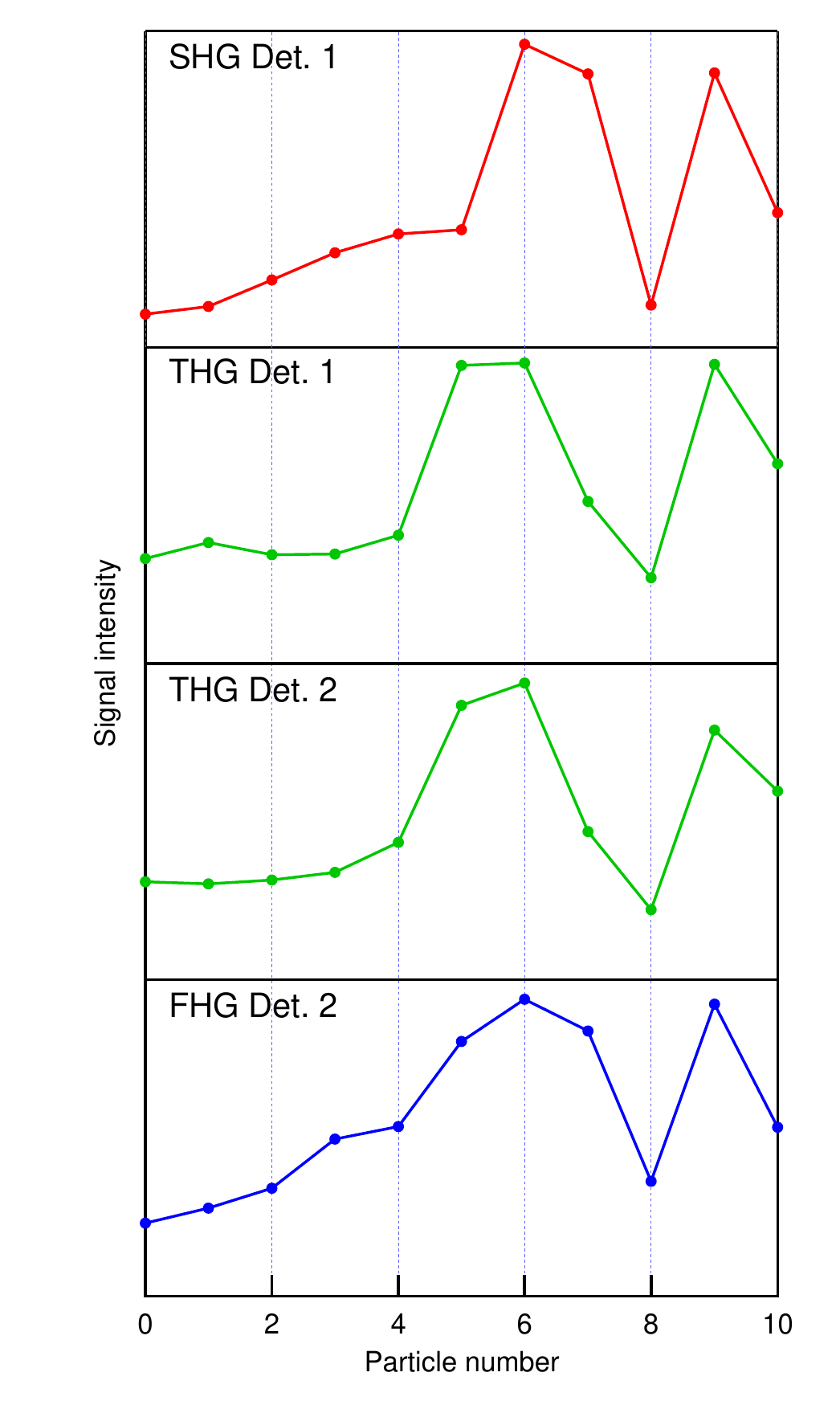}
\caption{Forward detected signals at the different harmonics generated by 10 distinct HNPs on a microscopy substrate. SHG is measured by detector 1, FHG by detector 2 and THG by both independently.} \label{FigSI:4}
\end{figure}

Intensity ratio measurements by individual BFO HNPs were performed using two different detectors to minimize the need of efficiency corrections among different data sets. As reported in Fig.~\ref{FigSI:4}, SHG and THG were measured by detector 1 (SPD-A-VISNIR ultra-low-noise single photon counting module, Aurea Technology) and THG and FHG by detector 2 (H7732-01 low noise sideon
photomultiplier tube, Hamamtsu). The traces highlight the particle-to-particle signal intensity variations, which come from differences in sizes (all signals are expected to scale as the particle volume squared), orientations, and possibly varying radiation patterns. We further confirmed these results on magnitude estimation among the different nonlinear orders employing a modified set-up with a NA 0.4 reflective Al-coated objective in the forward arm (Newport) and  detecting all harmonics by an EM-CCD (Andor, Ixon3) placed at the imaging output of the spectrometer.

\subsection{Ensemble measurements on BFO particle pellets}

\begin{figure}
\centering
\includegraphics[width=7cm, keepaspectratio]{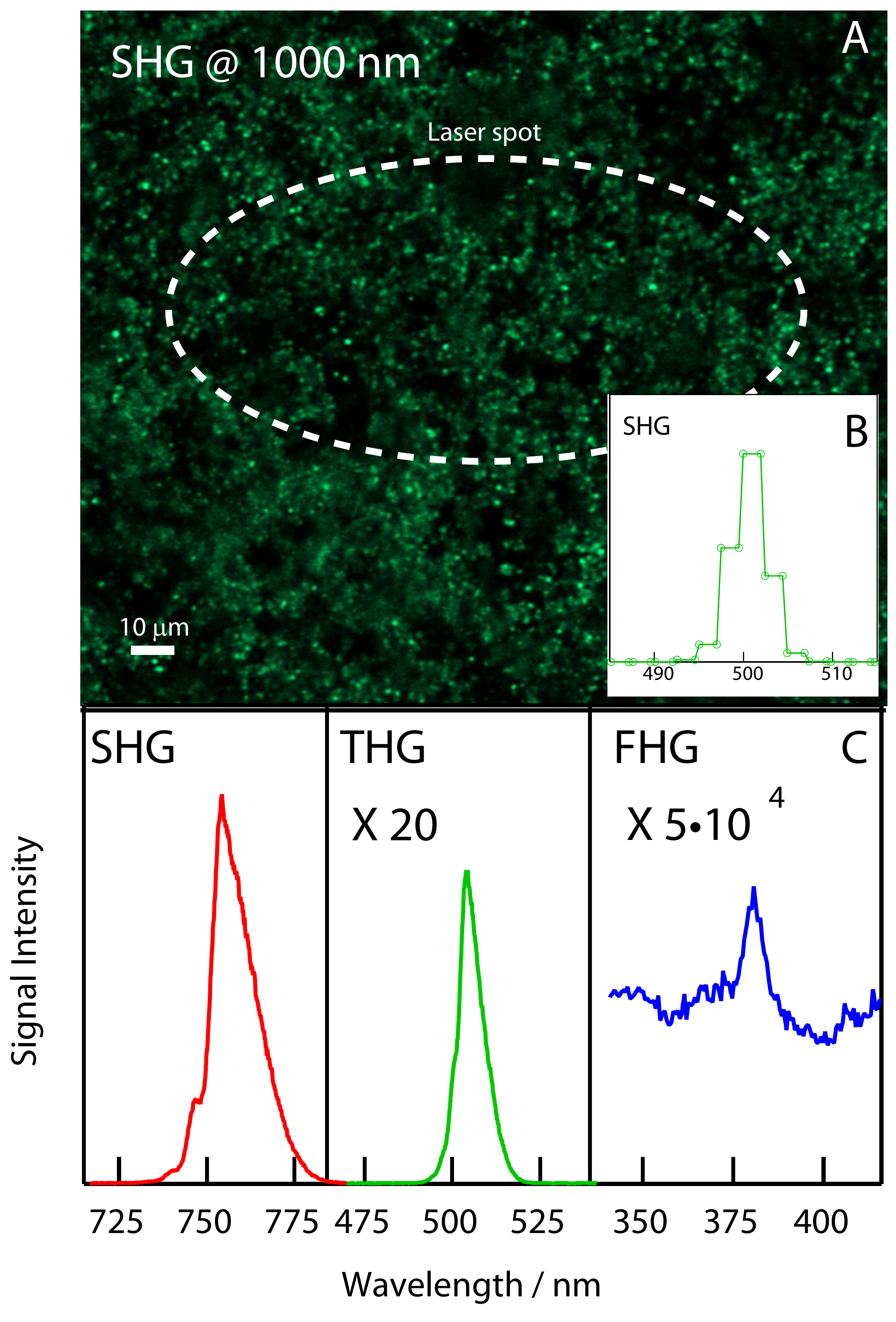} 
\caption{A. SHG image of the surface of BFO HNPs pellet obtained at 1$\mu$m excitation. B. SHG spectrum associated to image in A.  C. Harmonic spectra generated by the BFO pellet using the KHz laser system tuned at 1.5 $\mu$m. The relative intensities are corrected for exposition time and spectral properties of the optical components. The dashed line in A indicates the dimension of the focal spot on the sample (at $\frac{1}{e^2}$) taking into account the 60$^{\circ}$ beam incidence.} \label{FigSI:5}
\end{figure}

\begin{figure}
\centering
\includegraphics[width=10cm, keepaspectratio]{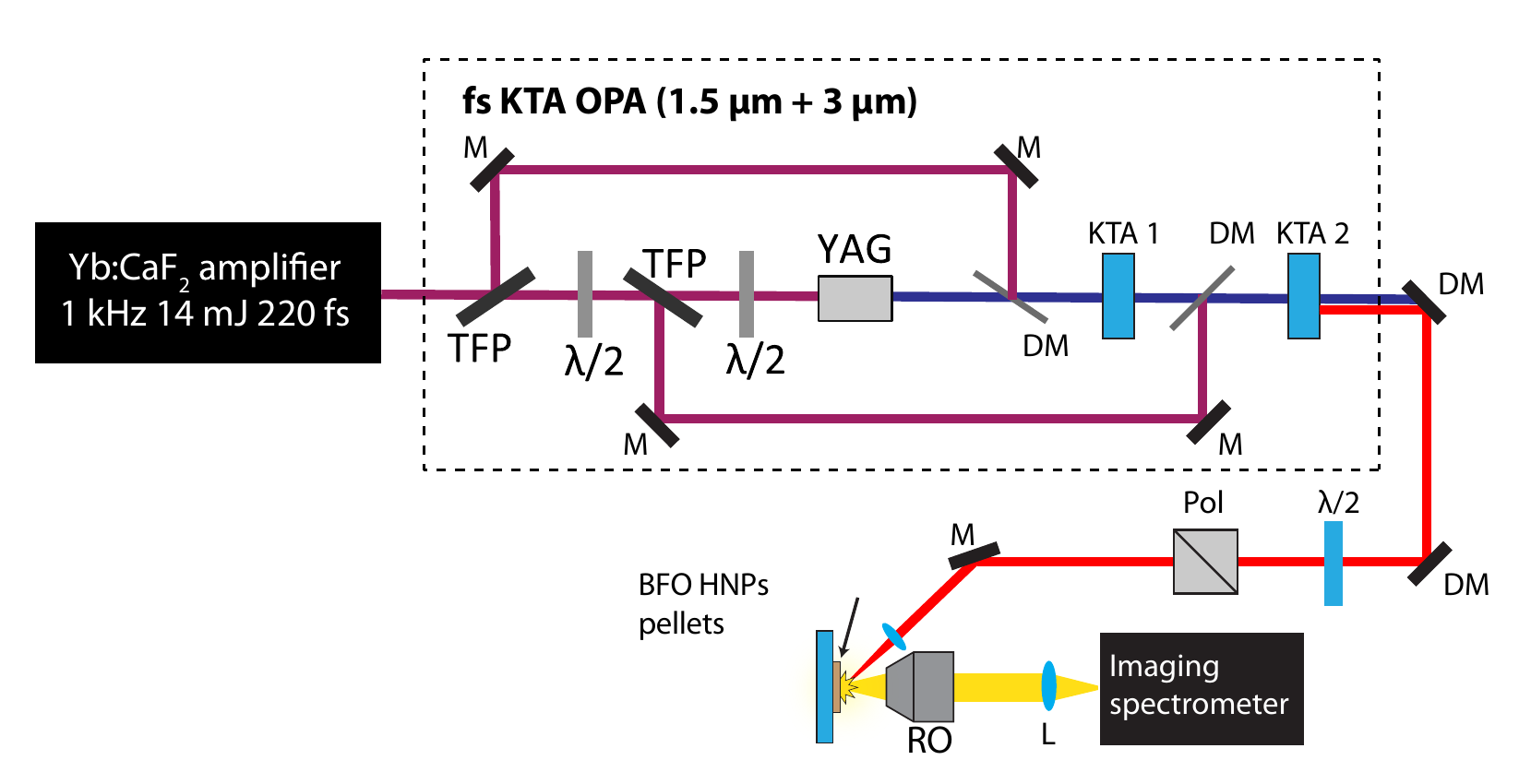}
\caption{Experimental setup for  harmonic generation by BFO HNPs pellets using $\mu$J energy pulses from a femtosecond parametric amplifier. \textit{DM}: dichroic mirror, \textit{TFP}: thin film polarizer, \textit{RO}: reflective objective.} \label{FigSI:6}
\end{figure}

In Fig.~\ref{FigSI:5}A, we provide a SHG image of the BFO HNPs pellet surface obtained by a commercial multiphoton microscope (Nikon A1R-MP) coupled with a Ti:sapphire oscillator (Mai Tai Spectra Physics). The epi-collected signal was processed by a Nikon A1 descanned spectrometer. The image scale bar is 10 $\mu$m. One can see how the SHG intensity of  HNPs is modulated by their diverse orientation and that most of the particles appear as bright diffraction limited spots. The emission spectrum averaged over the whole image is reported in Fig.~\ref{FigSI:5}B.

For comparing relative intensities of the harmonics on  BFO HNPs on dry pellets we relied on  the  laser set up reported in Fig.~\ref{FigSI:6}.   This system delivers  $\approx$80 fs pulses  at 1.5  $\upmu$m  generated in an OPA pumped by a 1 kHz 14 mJ 200-fs Yb:CaF2 CPA laser. The OPA is based on KTA crystals and seeded by a supercontinuum generated in a bulk YAG plate and delivers ~ 1.5 mJ signal pulses. The signal beam is filtered out at the OPA output using a set of dichroic mirrors, the energy is attenuated using a half-wave plate and a polarizer and then focused onto the sample using $f$=200 mm CaF$_2$ lens at 60$^{\circ}$ incidence. The  harmonic signals are collected in  reflection geometry using a Schwarzschild objective (\textit{ReflX}, Edmund Optics),  imaged onto the slit of a  imaging spectrometer, and detected using an  EM-CCD (Andor, Ixon3). In Fig.~\ref{FigSI:4}C, we present the spectra of the different harmonic generated by the pellet. The relative intensities are corrected for CCD exposure time and spectral sensitivity and for grating efficiency and can be quantitatively compared. 

\clearpage

\providecommand*\mcitethebibliography{\thebibliography}
\csname @ifundefined\endcsname{endmcitethebibliography}
  {\let\endmcitethebibliography\endthebibliography}{}

\end{document}